# Artificial General Intelligence (AGI) for the oil and gas industry: a review

Jimmy Xuekai Li, Tiancheng Zhang, Yiran Zhu, and Zhongwei Chen*

*School of Mechanical and Mining Engineering, The University of Queensland, Brisbane, QLD 4072, Australia*
*Corresponding Author: zhongwei.chen@uq.edu.au*

**Abstract**

Artificial General Intelligence (AGI) is set to profoundly impact the oil and gas industry by introducing unprecedented efficiencies and innovations. This paper explores AGI's foundational principles and its transformative applications, particularly focusing on the advancements brought about by large language models (LLMs) and extensive computer vision systems in the upstream sectors of the industry. The integration of Artificial Intelligence (AI) has already begun reshaping the oil and gas landscape, offering enhancements in production optimization, downtime reduction, safety improvements, and advancements in exploration and drilling techniques. These technologies streamline logistics, minimize maintenance costs, automate monotonous tasks, refine decision-making processes, foster team collaboration, and amplify profitability through error reduction and actionable insights extraction. Despite these advancements, the deployment of AI technologies faces challenges, including the necessity for skilled professionals for implementation and the limitations of model training on constrained datasets, which affects the models' adaptability across different contexts. The advent of generative AI, exemplified by innovations like ChatGPT and the Segment Anything Model (SAM), heralds a new era of high-density innovation. These developments highlight a shift towards natural language interfaces and domain-knowledge-driven AI, promising more accessible and tailored solutions for the oil and gas industry. This review articulates the vast potential AGI holds for tackling complex operational challenges within the upstream oil and gas industry, requiring near-human levels of intelligence. We discussed the promising applications, the hurdles of large-scale AGI model deployment, and the necessity for domain-specific knowledge in maximizing the benefits of these technologies.

# 1. Introduction

The oil and gas industry, a cornerstone of the global economy, is navigating a landscape marked by transformative opportunities and unprecedented technical challenges (Prestidge, 2022). As the industry undergoes a pivotal shift towards sustainable energy sources, it faces mounting pressure to enhance efficiency, reduce environmental impact, and fulfill the growing energy demands of a global population (Hassan et al., 2024; Zohuri, 2023). This transition requires navigating complex, high-risk operations across some of the planet's most challenging environments, encompassing the exploration, development, production, storage, and abandonment (Epelle and Gerogiorgis, 2020; Fagley, 2014; Islam and Khan, 2013; Speight, 2014).

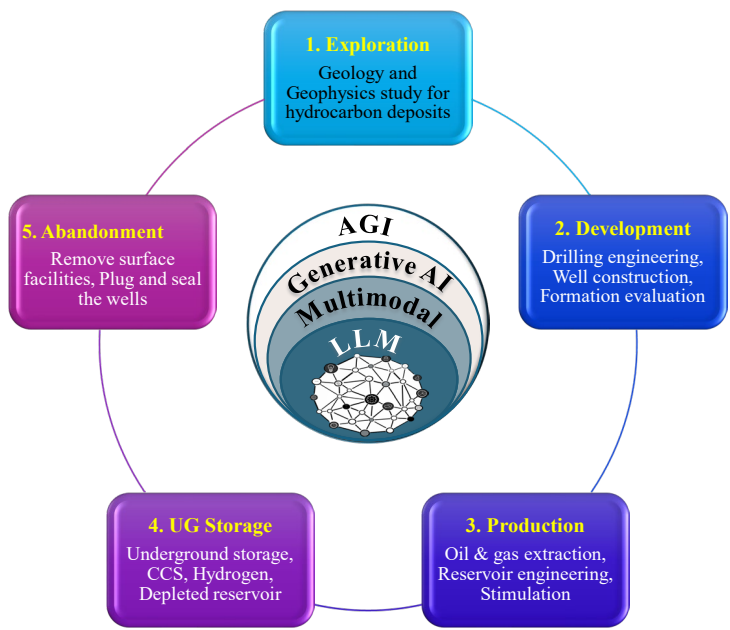

**Figure 1. Lifecycle and the critical stages of the upstream oil and gas industry. Artificial General Intelligence (AGI) and its subsets generative AI, multimodal and especially the large language model (LLM) across each stage would enhance efficiency and innovation in hydrocarbon exploration & production and underground storage.**

A general lifecycle of the oil and gas industry (Hussein, 2023; Tatjana, 2017) can be described by **Figure 1**. It delineates the key stages in the upstream sector of the industry. The initial exploration phase demands not only a deep understanding of geological structures but also the ability to accurately interpret subsurface data (Tearpock and Bischke, 2002; Wellmann and Caumon, 2018). Following exploration, the appraisal phase seeks to delineate the hydrocarbon reservoir's potential (Amado, 2013). Development operations then take center stage, involving drilling engineering, well construction, and comprehensive formation evaluation (Darling, 2005). Next, the production phase deploys reservoir engineering techniques and advanced stimulation methods to optimize and maximize the recovery (Alvarado and Manrique, 2010). This phase is pivotal in effectively managing the field's productive lifespan and extending the viability of wells. At post-production, the focus might shift to innovative storage solutions depending on the reservoir condition, reflecting the industry's commitment to resource management and environmental responsibility. Options range from underground formations for hydrocarbon storage, such as gas storage facilities (Tarkowski, 2019), to carbon capture and storage (CCS) (Bui et al., 2018) and hydrogen storage in depleted reservoirs (Muhammed et al., 2023). In the final stages of a field's life, abandonment protocols are enacted where safety and environmental restoration take precedence. This involves the careful removal of surface infrastructure and the sealing of wells, underscoring the industry's dedication to sustainable practices (Babaleye et al., 2019).

In this complex landscape, the adoption of Artificial Intelligence (AI) has become more prevalent in each stage of the oil and gas industry. These technologies have been used at the forefront of tackling the industry's challenges, offering innovative solutions across the value chain. Some comprehensive review and summaries of the task specific or narrow AI applications in the oil and gas industry can be found in these literatures (Hajizadeh, 2019; Pandey et al., 2020; Sircar et al., 2021; Tariq et al., 2021; Waqar et al., 2023; Zhong et al., 2022).

The role of AI in this context has been proving crucial and cannot be overstated. This leads people to have more hope for the on-going development of a new form of generalized AI or the Artificial General Intelligence (AGI) for the oil and gas industry. From enhancing exploration success rates and drilling efficiency to advancing safety and sustainability, AI and AGI are reshaping the future of oil and gas operations (Jacobs, 2024; Ma et al., 2023; Marlot et al., 2023; Ogundare et al., 2023; Paroha and Chotrani, 2024; Waheed, 2023; Wang et al., 2023; Weijermars et al., 2023). As this sector evolves in response to the shifting energy landscape, it leverages AI to not only address technical complexities but also to pioneer a path toward a more sustainable and efficient future. This integration of cutting-edge technology is a testament to the industry's resilience and forward-thinking ethos, promising to drive progress and innovation in the years to come.

The rest of this paper is organised as follows. Section 2 discuss the foundations of AGI, including its history and underlying technologies, such as LLMs, computer vision, and multimodal AI. Subsequent sections thoroughly examine the deployment of these technologies within the industry, focusing on applications in data analysis, geophysics interpretation, operational optimization, and beyond. The review highlights the latest innovative AGI applications that are enhancing exploration, safety, and collaboration while also addressing the accompanying challenges and workforce implications. It further outlook the future perspective of AGI in the oil and gas industry. The conclusion synthesizes the review's main findings, emphasizing AGI's strategic importance and potential for addressing complex operational challenges in the oil and gas industry.

Table 1. Comparison of Narrow AI and AGI.

| Aspect | Narrow AI | AGI |
|---|---|---|
| **Definition and Scope** | Designed to perform specific tasks within a limited range or context, such as classification or image recognition systems. | General cognitive abilities, can understand, learn, and apply intelligence to solve any problem across a wide range of domains. |
| **Learning and Adaptability** | Learns from large amounts of labeled data within its narrow scope, limited adaptability to its defined domain. | Designed to learn and reason in various domains, adaptable to new scenarios, potentially with less data. |
| **Complexity and Development** | Less complex, widely used in various industries, involves machine learning and rule-based systems. | Far more complex, theoretical, and experimental, aims to replicate human brain processing. |
| **Consciousness and Self-Awareness** | No self-awareness or consciousness, operates purely within algorithmic constraints. | Theories often suggest potential for self-awareness, although this is a subject of debate. |
| **Applications** | Applications are vast but specific, such as spam filtering, recommendation systems, and autonomous vehicles. | Applications could be revolutionary, ranging from running corporations to conducting scientific research. |
| **Challenges** | Data bias, need for large data sets, inability to generalize beyond trained tasks. | Ethical considerations, the control problem, technical difficulty in creating AGI systems. |
| **Impact** | Substantial impact in optimizing and automating tasks, driving innovation within its scope. | Could revolutionize every aspect of human life but poses risks and uncertainties. |

## 2. Artificial General Intelligence (AGI) Background

AGI is an area of research and innovation aimed at developing computational systems capable of understanding, learning, and performing any intellectual task that a human being can (Goertzel, 2014; Obaid, 2023; Youvan, 2024). Unlike narrow or weak AI, which is designed to excel at a specific task or set of tasks, AGI seeks to achieve a versatile and adaptive form of intelligence that mirrors human cognitive abilities. This includes the capacity for reasoning, problem-solving, perception, and abstract thinking across diverse domains without needing task-specific programming. **Table 1** provides a simple comparison of the conventional narrow AI and the AGI based on summaries of multiple literatures (Güner, 2022; Hoque, 2023; Kanade, 2022; Koc, 2023; VK, 2022).

The advancement of AI in the last decade, especially the recent couple of years (**Figure 2**), is an unprecedent phenomenon primarily driven by the convergence of three key factors: the burgeoning availability of diverse and voluminous training datasets (Pal et al., 2023), the evolution and enhancement of AI infrastructures (Gill et al., 2019; Jeon et al., 2021), and the innovation within AI modelling techniques, notably Generative Adversarial Networks (GANs) (Creswell et al., 2018; Goodfellow et al., 2014, 2020), diffusion models (Borji, 2022; Croitoru et al., 2023; Kingma et al., 2021), and Transformers (Khan et al., 2022; Lin et al., 2022; Vaswani et al., 2017). These elements have all together propelled AI capabilities forward, setting the stage for the rise of AGI. The discourse around AGI has permeated various sectors, including agriculture (Lu et al., 2023), education (Latif et al., 2023), healthcare (Li et al., 2023b; Liu et al., 2023a), and Internet of Things (IoT) (Dou et al., 2023), with its popularity surging post the introduction of OpenAI's ChatGPT (Wu et al., 2023b). It's crucial to recognize that the ascendancy of AGI is not achieved by a single entity but an outcome of collective work across academia and industry. A landmark in AGI's journey was the unveiling of the Transformer architecture by Google in 2017 (Vaswani et al., 2017), a model designed to replace Recurrent Neural

Networks (RNNs) with self-attention mechanisms enabling sequential processing.

AGI's narrative took a dramatic turn with the advent of BERT (Bidirectional Encoder Representations from Transformers) (Devlin et al., 2018), which, through its revolutionary pre-training on extensive corpora followed by fine-tuning on specific tasks, set new benchmarks in natural language processing (NLP). The advent of BERT heralded the dawn of foundation models (Koroteev, 2021), which shift the research focus from task-specific training to the development of large-scale, pre-trained models that adapt to various applications through subsequent fine-tuning, few-shot, or zero-shot learning strategies. This new wave has not only cemented the role of large-scale, task-agnostic models, conventionally known as foundation models, within AGI but has also underscored their significance in domains beyond NLP, like reinforcement learning (RL) (Wang et al., 2020) and computer vision (CV) (Khan et al., 2021).

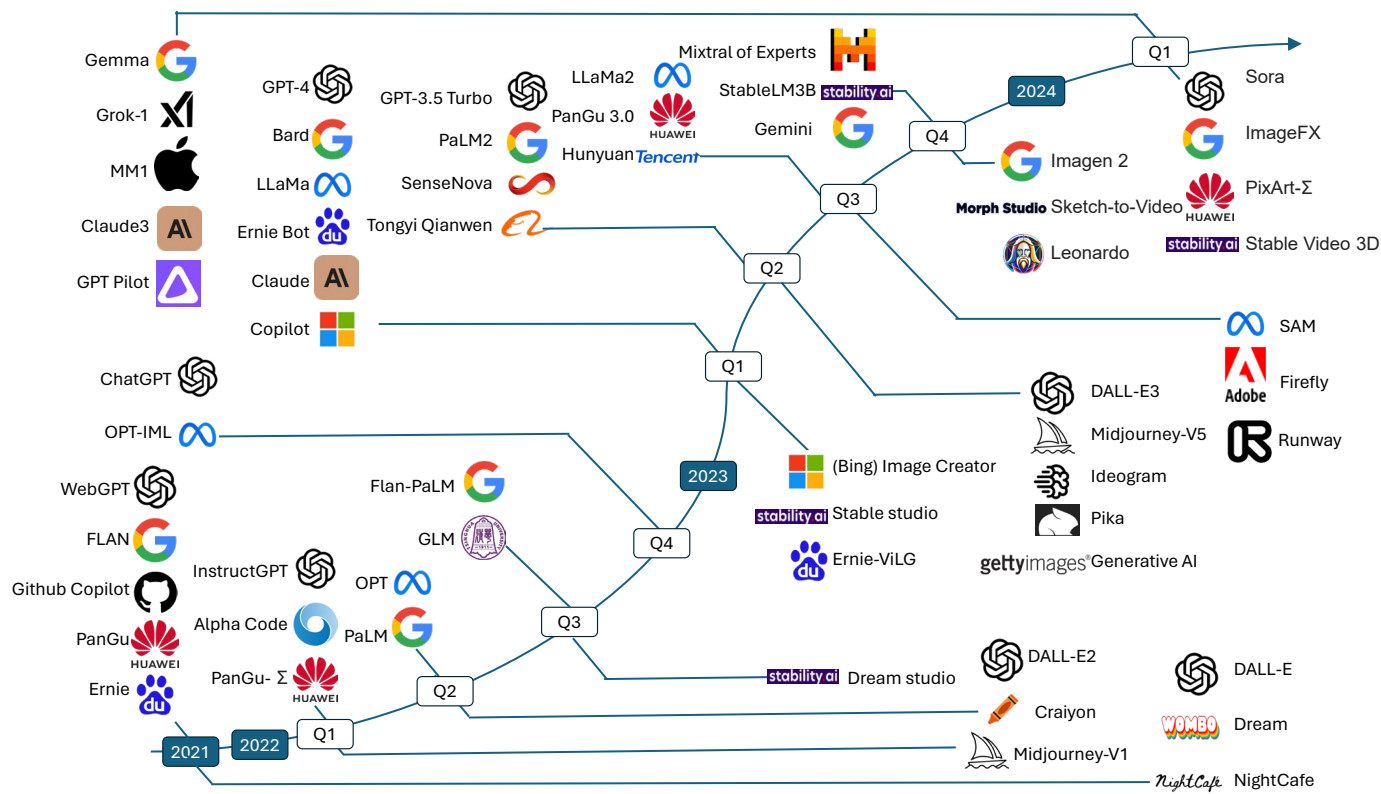

Figure 2. Recent developments of large language models (LLM, left side of the timeline curve) and multimodal AI models (right side of the timeline curve), which combine the capabilities of LLM and CV (image, video, or 3D objects). Note: this figure does not intend to provide an exhaustive review or include all existing LLMs and multimodal models. Instead, it aims to highlight some of the most renowned or commonly used LLMs and multimodal models, which may also encompass APIs or web applications. The creation of this figure draws inspiration from the Figure 3 in Zhao et al., (2023a), serving as a reference point rather than a comprehensive catalogue.

For instance, visual foundation models, typified by DALL·E (Offert and Phan, 2022) and Imagen (DeepMind, 2023; Gautam et al., 2024) and Sora (OpenAI, 2024), exemplify the migration of AGI into the visual domain, showcasing adaptability across a multitude of tasks ranging from style transfer, image editing to image and video generation.

The AGI ecosystem is rapidly evolving to embrace multimodal foundation models capable of handling diverse data forms, including text, images, video, and audio. Such models, exemplified by ChatGPT and GPT-4, facilitate cross-modal knowledge transfer and unlock capabilities for multifaceted tasks requiring holistic data interpretation (Brin et al., 2023; Egli, 2023).

Recent advances in large language models, like open source Grok-1 developed by X AI (XAI, 2024) trained with 314 billion parameters, have shown glimpses of AGI's potential, demonstrating a broad of understanding and reasoning that approximates a more generalized form of intelligence. As such, these models are increasingly regarded as next generation of mega scale of AGI, paving the way for an era where AI's applications are as equivalent as the human intellect.

## 3. Towards AGI paradigm for the oil and gas industry

The journey towards embedding AGI in the oil and gas industry signifies a shift from conventional operational frameworks to more intelligent, efficient, and sustainable practices. This section reviews the recent key applications of LLM, LLM-based agents and multimodal, outlines the pathway toward realizing an AGI paradigm in the oil and gas industry, highlighting strategic initiatives, technological advancements, and collaboration opportunities that pave the way for this ambitious integration.

Adopting AGI in the oil and gas industry involves leveraging advanced AI capabilities to achieve comprehensive cognitive functions enabling systems to perform a wide range of tasks autonomously (Weijermars et al., 2023). From exploration and production to refining and distribution, AGI can offer unparalleled insights, predictive analytics, and decision-making processes, leading to optimized operations and reduced environmental impact. However, transitioning to an AGI-driven framework requires addressing considerable hurdles, including technological readiness, data governance, workforce adaptation, and regulatory compliance.

## 3.1. Finetune LLM on domain knowledge

Although LLMs, such as GPT-4, LLaMA2, and Grok-1 models, have been trained through sophisticated methodologies that leverage vast amounts of data and computational resources, finetuning LLMs on domain knowledge and private data is a strategic approach to enhancing their utility, specificity, and effectiveness across various applications. It not only ensures that the outputs are highly relevant and tailored to specific needs but also addresses concerns related to data privacy, bias, and real-world applicability.

In the realm of geoscience, Deng et al., (2023) pioneers by introducing K2, the inaugural LLM specifically tailored for geoscience, accompanied by an innovative suite of resources aimed at enhancing LLM research within this field. The creation of K2 marks a major stride toward bridging the gap between advanced LLM capabilities and the intricate requirements of geoscience research and application. This endeavour is supported by the development of GeoSignal, a novel instruction-tuning dataset, including Natural Instructions (Mishra et al., 2021), the AI2 Reasoning Challenge (Clark et al., 2018), Stanford Alpaca (Taori et al., 2023), and Dolly-15k (Conover et al., 2023), designed to refine LLM responses to geoscience queries, and GeoBench, a benchmark for assessing LLM performance in geoscience contexts.

Deng et al., (2023) proposes a comprehensive adaptation strategy for pre-trained general-domain LLMs, i.e., LLaMA-7B (Meta, 2023), to the geoscience domain, utilizing over 5.5 billion tokens from geoscience texts for further training and fine-tuning the model with GeoSignal's expert-curated data (**Figure 3**). This approach not only amplifies the model's proficiency in geoscience knowledge but also equips K2 with tool-using capabilities, thus paving the way for a multitude of geoscience applications. Through rigorous evaluations on GeoBench, the paper demonstrates the model's enhanced understanding and utility in geoscience, setting a precedent for future research and applications in domain-specific LLMs. The open-source release of K2 and its datasets underscores the collaborative spirit of this research, inviting broader participation in the exploration and expansion of AGI in specialized geoscience domains.

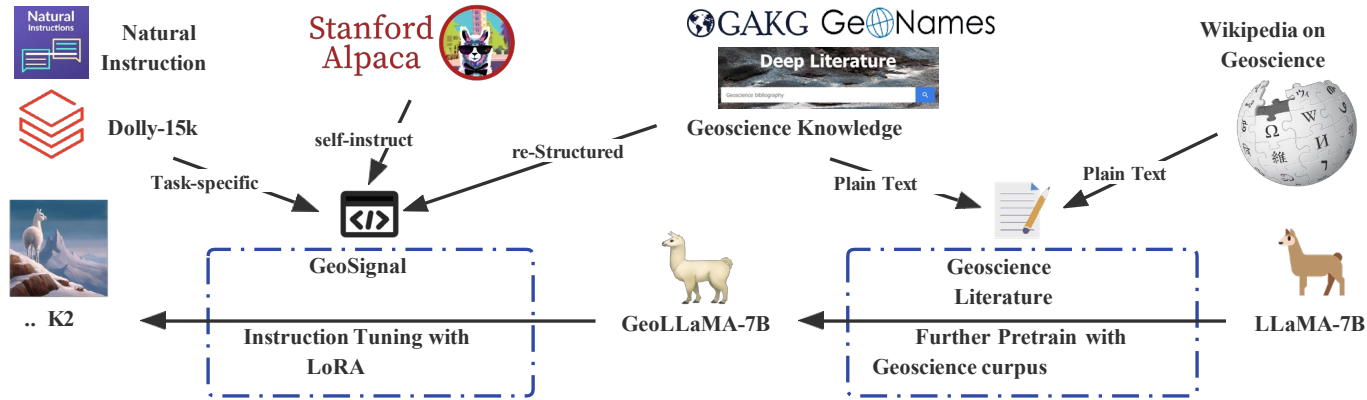

Figure 3. The training pipeline for K2 comprises two key stages: the first involves further pre-training to integrate geoscience knowledge, enhancing the model's expertise in this domain. The second stage, instruction tuning with low-rank adaptation (LoRA) (Hu et al., 2021), is designed to fine-tune the model for human-like interaction. This involves adapting the model to follow human instructions, emulate human responses, and align more closely with human conversational patterns, ensuring a more natural and intuitive user experience (After Deng et al., (2023)).

As a language model, K2 is capable of understanding geoscience materials and tailoring responses with suitable prompts. As a generative model, K2 excels in creating detailed paragraphs and responding to queries with precise answers, functioning similarly to a dynamic knowledge base that offers geoscientists professional assistance. Furthermore, through the use of tools like Augment K2, the user can incorporate external information and functionalities to produce reliable and effective outcomes. For instance, when posed with a question, K2 can initiate a search using the GAKG's API, update its actions based on new data, and craft a prompt that leads to generating the most appropriate answers. **Figure 4** illustrates this process in action.

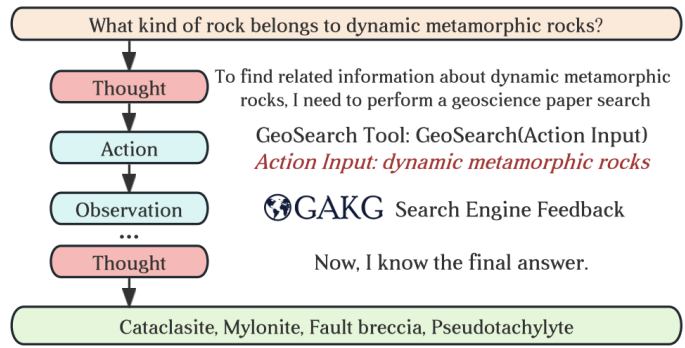

**Figure 4. An example of tool augmented K2 for geoscience Q&A (After Deng et al., (2023)).**

While K2 is optimized for geoscience, making it highly effective within this domain, K2's and its base model LLaMA-7B's small scale (7 billion parameters in LLaMA-7B) limit its versatility compared to the more advanced LLM models such as GPT-4, which boasts 1.76 trillion parameters and a Mixture of Experts (MoE) architecture (Betts, 2023; Shazeer et al., 2017).

The MoE architecture is a scalable machine learning approach that dynamically allocates computational resources across different "expert" models to efficiently process diverse tasks and data types, enhancing model performance and adaptability. Thus, we suggest in the future work it should consider scaling up the model size for increased complexity handling, incorporating a MoE architecture for improved efficiency and adaptability, and expanding domain-specific training datasets with the latest geoscience data.

### 3.2. Enhancing Equipment Maintenance with LLM

The equipment maintenance strategies within the oil and gas sector have been enhanced through the application of LLMs such as GPT-3 (Brown et al., 2020) and BERT (Devlin et al., 2018). LLMs analyse vast datasets, encompassing operational parameters, historical performance data, and environmental factors. By understanding patterns and predicting potential failures, these models enable a shift from traditional reactive maintenance to a more efficient preventive approach. This not only helps in scheduling maintenance activities before the actual occurrence of failures but also dramatically reduces unplanned downtimes and associated costs. Recognizing the sector's unique challenges, including the management of vast unstructured data and industry-specific terminology, Abijith et al., (2023) tailors these advanced NLP technologies to better understand and process maintenance-related texts. This initiative seeks to unlock valuable insights from documents like manuals, work orders and technical records (**Table 2**), which are crucial for optimizing both preventive and corrective maintenance operations and decision-making processes. Central to their approach is the development of a custom LLM based on RoBERTa, a model chosen for its proven effectiveness in various NLP tasks. The customization process involves extensive training on a corpus specifically gathered for this purpose, including equipment manuals and plant maintenance records, to ensure the model's familiarity with the industry's specific language. This process also includes the creation of a custom tokenizer, essential for accurately segmenting text into analysable units, thus enabling the model to effectively decode the specialized vocabulary of the oil and gas industry.

**Table 2. Distribution of data used for training and tokenization in the custom LLM for equipment maintenance (Abijith et al., 2023).**

| Data Set | Number of words |
|---|---|
| Manuals | 7.865 Million |
| Materials Data | 1 Million |
| Notification Data | 130 Million |
| Work Order Data | 300 Million |
| Work Task Data | 8.3 Million |

It details the rigorous training regime, employing data parallel and distributed data parallel techniques to enhance training efficiency and scalability (Zhao et al., 2023b). The model's performance is thoroughly evaluated through tasks designed to reflect real-world applications, such as equipment segregation and masked word predictions. Despite facing challenges, particularly in handling industry-specific abbreviations and terms, the findings illuminate the path for considerable improvements, emphasizing the importance of comprehensive, domain-focused training datasets. By suggesting ways to overcome identified limitations and proposing areas for further exploration, such as refining the tokenization process and expanding the training datasets, the study encourages for a continued effort towards integrating advanced NLP technologies into the sector. The proposed future work aims not only to refine the model's performance but also to explore its application across a wider range of maintenance tasks, expected to markedly elevate the efficiency and accuracy of maintenance strategies in the oil and gas industry.

Based on similar approach, Paroha and Chotrani, (2024) finetuned two advanced LLMs (i.e., TimeGPT and Time-LLM), and applied them in predicting maintenance needs for Electrical Submersible Pumps (ESPs) for the oil production. It was found that both models were effective in identifying key indicators of ESP health, which aligns with established industry knowledge. TimeGPT (**Figure 5**) demonstrated slightly better performance overall, achieving higher accuracy, precision, recall, and AUC-ROC values compared to Time-LLM. This demonstrates the potential of these AI models to enhance predictive maintenance strategies within the sector. However, the integration of these AI models into existing industrial setups poses substantial challenges, including the necessity for high-quality data and compatibility with existing systems.

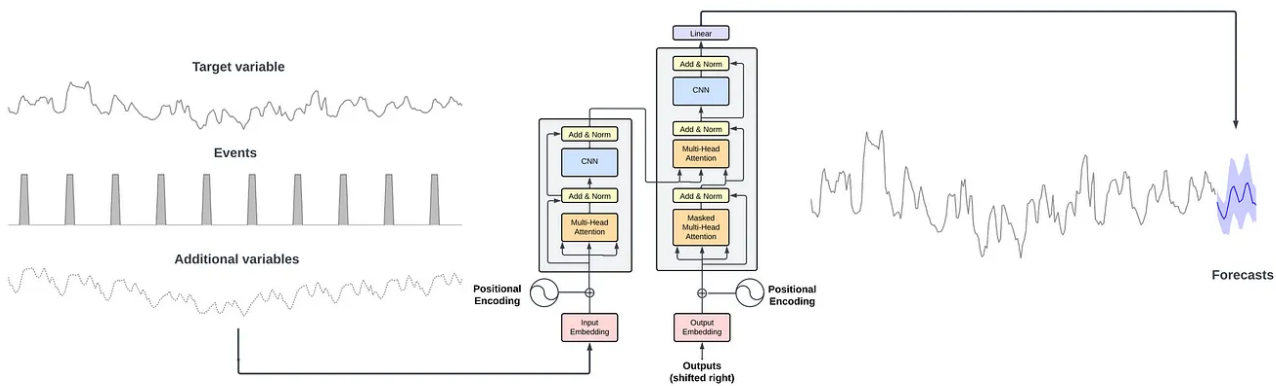

**Figure 5. The architecture diagram of TimeGPT, (After Garza and Mergenthaler-Canseco, (2023)).**

### 3.3. LLM as coding tool for oil and gas industry

LLMs, as part of generative AI, play a critical role in moving towards AGI by leveraging their ability to understand and generate human-like text to assist with programming tasks, automation, and especially generating executable codes based on natural language prompts so as to enhance various facets of software development, data handling, and operational efficiency. These models are particularly adept at automating and improving tasks that traditionally require considerable human intervention and expertise.

In the area of geosciences, Weijermars et al., (2023) recently integrate the LLM (ChatGPT) into the coding and data processing of full-waveform inversion (FWI) for the seismic data, demonstrating generative AI's profound impact on handling complex geoscientific data. They input a prompt to ChatGPT asking for a Python function based on Weiner optimum filtering to perform spiking deconvolution on a series of seismic shot records. The AI efficiently returns a function that handles the input parameters (e.g., seismic shot gather, length of the spiking filter, and pre-whitening percentage) and produces the deconvolved seismic data (**Figure 6**). Rapid code generation by LLM reduces the time that engineers and researchers spend on coding, allowing them to focus more on analysis and interpretation of results. In addition, the generated code that is also syntactically correct and functionally effective, reducing the likelihood of human error in manual coding.

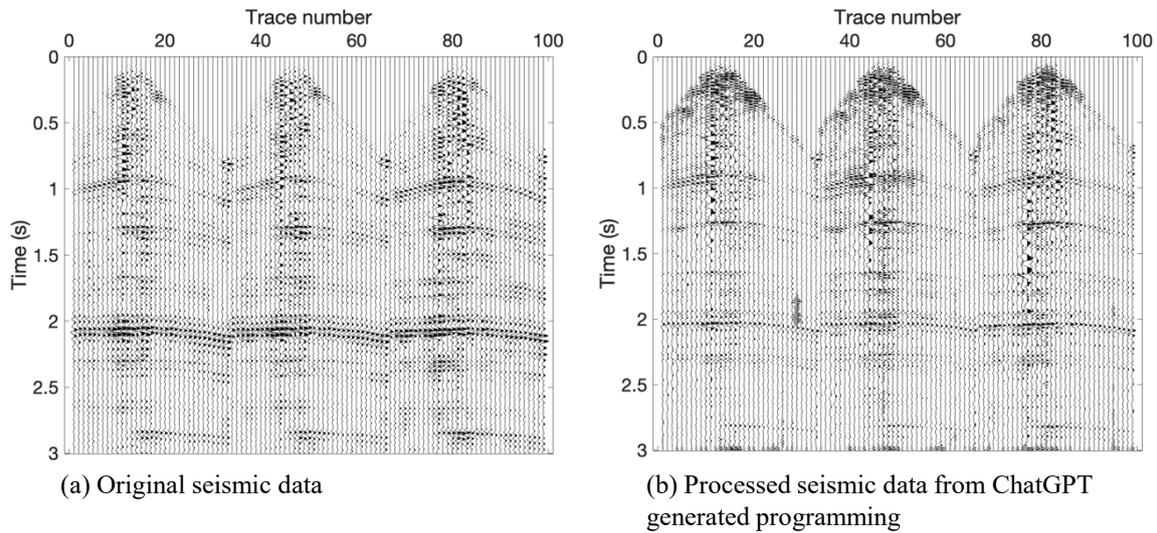

(a) Original seismic data     (b) Processed seismic data from ChatGPT generated programming

**Figure 6. Comparison of Seismic Shot Gathers from a 2D Landline in East Texas, USA Panel: (a)** shows the data before applying spiking deconvolution, while panel **(b)** displays the results after deconvolution. The Python code used for this process was generated by ChatGPT based on a prompt requesting a function for spiking deconvolution using Weiner optimum filtering. This function processes seismic shot gathers, with inputs including the gather itself, the length of the spiking filter, and the pre-whitening percentage. The output is the deconvolved seismic data, demonstrating the effectiveness of the filtering approach (after Weijermars et al., (2023)).

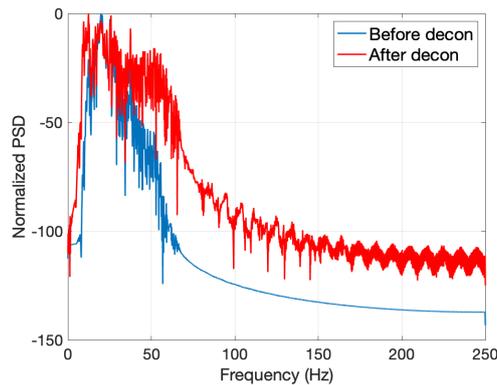

**Figure 7. Normalized power spectral density of the average trace before and after spiking deconvolution.** The spectral density is shown in blue prior to deconvolution and in red after the process. This comparison confirms an enhancement in resolution following the application of spiking deconvolution code generated by ChatGPT (after Weijermars et al., (2023)).

**Figure 7** showcases the efficacy of the generated code through a graphical representation of the normalized power spectral density (PSD) before and after spiking deconvolution. The comparison confirms resolution enhancement, validating the LLM's capability in generating functional and reliable code that can be directly applied in geoscientific research. While LLMs can generate code quickly, the responsibility remains on the researcher to verify the accuracy and appropriateness of the output, especially in scenarios involving novel or complex data interpretations. Also depending on specific research needs, the generated code may require adjustments or enhancements to better align with unique project specifications or data characteristics.

Dhelie et al., (2023) also applied ChatGPT to oil and gas exploration tasks, highlighting successful implementations in seismic image processing during pre-stack processing stages such as denoising, deblending, deghosting, and debubbling. It emphasizes the ease of use and reductions in processing time achieved with AI tools such as ChatGPT compared to traditional CPU-intensive methods. **Figure 8** illustrates the use of the ChatGPT for generating programming code for seismic image enhancements, noting its simplicity and efficiency in producing usable code with minimal input.

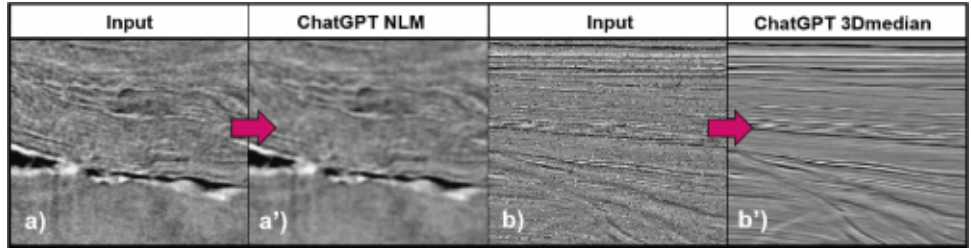

**Figure 8.** Seismic images enhancement by the programming code generated by the ChatGPT (After Dhelie et al., (2023)).

In sum, the use of generative AI in generating code for data processing in geosciences and all other sectors in the oil and gas industry represents a big leap forward in operation and research productivity and capability. As AGI technologies continue to evolve, their integration into petroleum engineering and geoscientific research is expected to deepen, offering even more sophisticated tools for data analysis and interpretation.

### 3.4. Multimodal AI on enhancing real-time production prediction

Multimodal AI refers to the models that integrate and process two or more different types of data modalities (such as text and images), reflecting an important step towards AGI. These systems are designed to leverage the complementary information available from different modalities to improve the accuracy and robustness of AI applications. Its ability to integrate diverse data types mirrors the versatility and adaptability required for AGI, marking a great advancement over narrow AI systems.

The upstream oil and gas industry are characterized by the vast and varied types of data it generates, including seismic data, drilling logs, core samples, and production data among others. This diversity of data spans numerous modalities, such as numerical data, text, and images, making multimodal AI particularly important for this sector. One noteworthy development is the integration of multi-modal data into predictive models for real-time oil production. Jiang et al., (2023) use an improved Artificial Fish Swarm Algorithm (AFSA) combined with a Long Short-Term Memory (LSTM) network to optimize production prediction models using multimodal data containing image features and production data.

The core of the proposed methodology is the fusion of different data modalities to improve the predictive accuracy of oil production models. The study uses image features extracted from indicator diagrams, which are traditional tools for monitoring equipment and production processes (**Figure 9**. These image features are combined with time-series production data, creating a rich, multimodal dataset that provides a comprehensive view of the production dynamics.

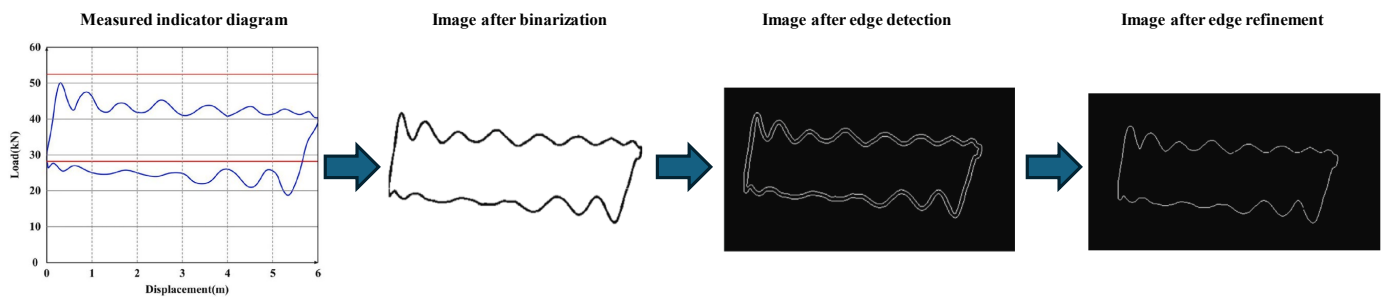

**Figure 9.** Feature extraction on indicator diagrams. The indicator diagram often contains rich information, e.g., load, displacement scale, coordinate axis, and so on, but it cannot be used directly for production prediction. The features extractions (after process e.g., binarization, edge detection and refinement) can be deemed as the features of the indicator diagram and used for production prediction (Modified from Li et al., (2024)).

The LSTM can process time series data effectively, while the AFSA is utilized to optimize the LSTM's hyper-parameters dynamically. This combination addresses two major challenges in traditional predictive modelling in the oil and gas industry: the reliance on single-modal data which limits predictive performance, and the manual setting of model parameters, which impedes achieving the optimal model configuration.

The effectiveness of the proposed model is demonstrated through experiments, which show that the AFSA-LSTM model achieves a mean absolute percentage error (MAPE) of just 4.313%, outperforming both traditional methods and other deep learning models. The integration of multimodal data not only enhanced the model's accuracy but also its robustness against varying operational conditions.

This study demonstrates the potential of multimodal AI applications in revolutionizing production prediction in the oil and gas industry. By integrating diverse data types and employing advanced machine learning techniques, operators can achieve more accurate predictions, leading to better resource management and optimized operational decisions.

## 3.5. Multimodal AI on enhancing standard knowledge extraction

The oil and gas industry have been generating huge amount of multimodal data during the complex and intensive operations, ranging from texts and documents to images and videos. Traditional methods, which often rely on manual processing of single-modal data, are inadequate for the semantic representation and integration of the industry's multimodal data. Multimodal AI could play a pivotal role to enhance the extraction and organization of knowledge from the vast and diverse data in the industry, which are often unstructured or semi-structured.

A recent study by Huang et al., (2024) explores an advanced approach for automatic extraction of multimodal knowledge specifically tailored for the standard documents in petroleum field. The algorithm designed to extract fragmented standard indicator attributes primarily involves two components: table extraction and paragraph extraction. Both modules rely on a part-of-speech classifier. Through table extraction, the user can derive ontology and indicator items directly from tables. Similarly, paragraph extraction allows the user to capture ontology from the applicable scope paragraphs in standard documents. The outputs from both extraction methods are then merged and incorporated into the indicator database. **Figure 10** presents a flowchart illustrating the process of the standard indicator extraction algorithm.

The new multimodal AI integrates various AI technologies, including text clustering, keyword extraction, and advanced classification techniques, to handle the multimodal nature of the data effectively. For instance, the study enhances text classification systems to automatically categorize documents and assigns them to predefined categories, facilitating easier management and retrieval of information. The approach not only utilizes traditional statistical methods but also incorporates rule-based classification technologies, allowing for dynamic adjustment according to specific industry needs.

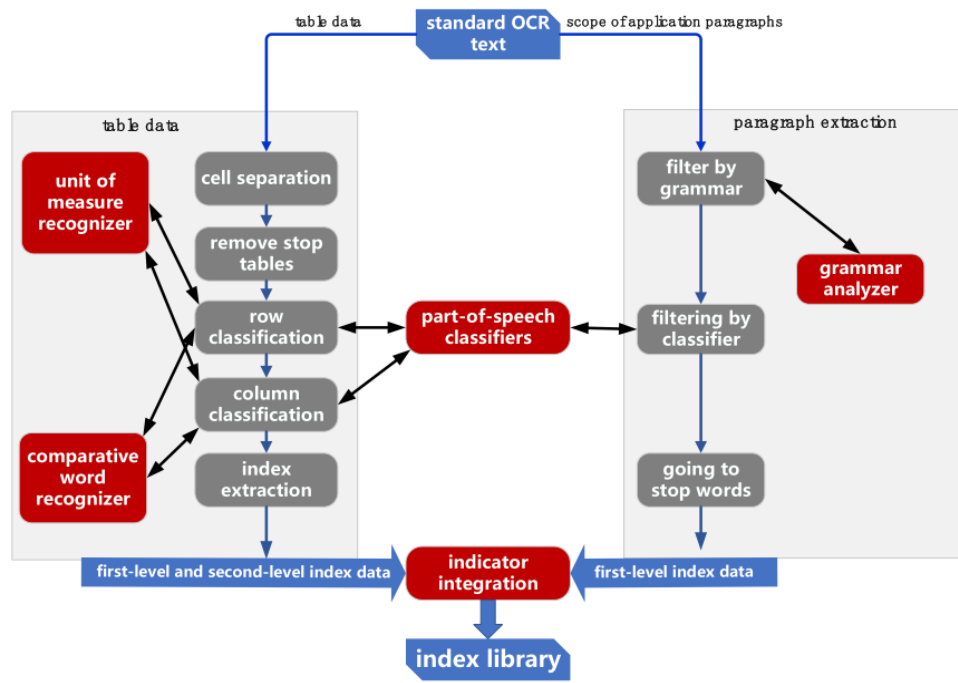

Figure 10. Flowchart of the standard indicator extraction algorithm (After Huang et al., (2024)).

The application of such multimodal AI systems enables the automatic organization and semantic analysis of vast quantities of data, which not only improves efficiency but also enhances the accuracy and accessibility of information. This ability to quickly and accurately extract and organize standard knowledge would also support the industry's ongoing efforts in innovation and regulatory compliance. By effectively leveraging the power of AI to process and analyse multimodal data, the industry can expect substantial advancements in how information is managed and utilized.

## 3.6. Multimodal AI on gas leakage detection

An important application of multimodal AI systems in the industry is in the robust sensor fusion, which refers to the process of combining sensory data from disparate sources to produce more accurate, consistent, and useful information than that provided by any individual sensor alone so as to capture a comprehensive understanding of the environment.

A recent study by Rahate et al., (2023) demonstrates the application of multimodal co-learning to evaluate the robustness of sensor fusion in gas detection systems, which is crucial for ensuring safety in various environments across the industry. The research focuses on using a combination of gas sensor data and thermal images to enhance the detection and identification processes (**Figure 11**).

It introduces a methodological innovation by employing multimodal co-learning, which can deal with scenarios where data from one or more sources may be noisy or unavailable. The multimodal approach integrates data from primary gas sensors and thermal cameras to form a robust detection system. This integration allows the system to compensate for missing or noisy data from one modality by leveraging information from the other, ensuring reliable operation under varying conditions. The experimental results confirm that multimodal sensor fusion can enhance the system's robustness compared to traditional single-source methods. For instance, the addition of low-resolution thermal modality data aids in compensating for up to 20% missing sensor data and 90% missing thermal image data. This capability is crucial in real-world industrial settings, where sensor outputs can be unpredictable and affected by environmental factors.

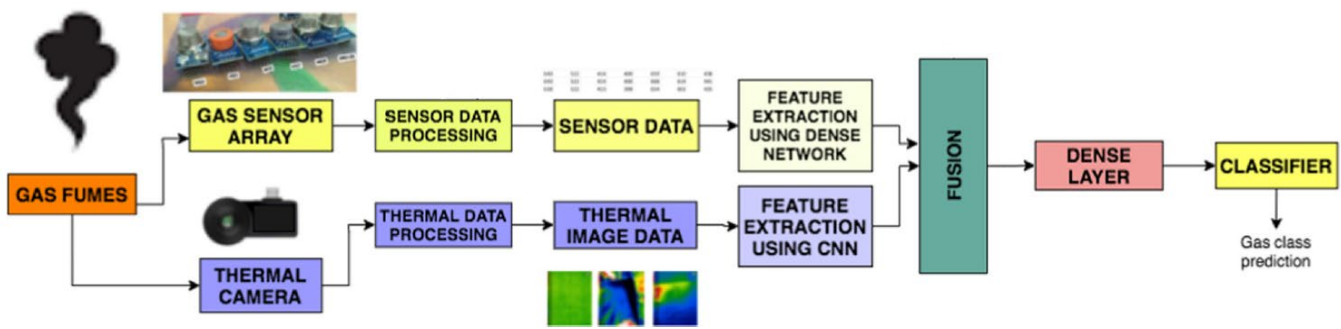

**Figure 11. Multimodal co-learning system architecture for evaluating the robustness of sensor fusion (After Rahate et al., (2023))**

In a later study, Attallah, (2023) introduces novel multimodal data fusion strategies termed "intermediate" and "multitask" fusion, which are more sophisticated compared to the multimodal co-learning focus in Rahate et al., (2023)'s work. Intermediate fusion involves integrating features at a mid-level using techniques like Discrete Wavelet Transform (DWT), improving the handling of spatial, spectral, and temporal information. Multitask fusion, on the other hand, employs Discrete Cosine Transform (DCT) for dimensionality reduction and feature integration from different AI architectures, providing a robust way to handle diverse data types efficiently. The system achieves high detection accuracies (98.47% and 99.25% for intermediate and multitask fusion respectively), which is a remarkable improvement over previous methodologies.

**3.7. Zero-shot learning on drill core images**

Zero-shot learning (ZSL) is a machine learning technique where a model learns to recognize objects, concepts, or classes that it has not seen during training. It relies on understanding general characteristics and properties that can be transferred from seen classes to unseen classes. This ability is a step forward towards creating more flexible and general-purpose models, which is one of the foundational goals in the pursuit of AGI.

Drill cores are invaluable in providing direct subsurface information and continuous record of the rock formation. The accurate measurement of fracture frequency or fracture density from drill core samples thus provides vital information that impacts the exploration and exploitation of natural resources, the assessment of geological hazards, and the design and construction of underground structures. The AI models have exhibited superior performance and efficiency compared to traditional manual core logging methods, offering an automated approach to geological and geotechnical analysis, which represents a technological shift aimed at enhancing the precision and efficiency of geological explorations. (Li et al., 2023a; Su et al., 2023). We tested different computer vision AI models on the core sample mask detection and classification task. **Figure 12** offers a comparative analysis of core sample instance segmentation with overlaid segmentation masks, applying three distinct AI models: Mask R-CNN, Mask2Former, and Segment Anything (SAM).

Mask R-CNN (He et al., 2017) relies on labelled data and utilizes transfer learning from pre-trained model to perform instance segmentation. Here, the precision of Mask R-CNN is evident as it can perform segmentation on the images of core samples with accuracy, but the requirement for extensive labelled datasets and the associated preparatory work limits its scalability and adaptability to new, unlabelled datasets. The more recent model Mask2Former (Cheng et al., 2022) refines the capabilities of its predecessor through the integration of transformer-based architectures, enabling a more sophisticated understanding of the global context within the image. Its performance, potentially surpassing Mask R-CNN, is indicative of advancements in AI, however, it still faces the challenges of data dependency and the need for a substantial annotation effort.

The latest Segment Anything (SAM) model (Kirillov et al., 2023) displays what is referred to as a 'zero-shot prediction', which does not require labelled data to predict and segment new instances. It operates on the principles of zero-shot learning, where the model generalizes from seen to unseen data, showcasing an approach that aligns with the trends towards AGI. This generalization capability is especially crucial in fields where labelling data is impractical due to the variety and complexity of instances, such as core samples with diverse geological features.

The generalization of SAM not only underscores the potential for AI models to progress towards AGI but also demonstrates a remarkable evolution in instance segmentation technology. The transition from reliance on labelled data to the ability to predict without prior direct examples illustrates a growing autonomy in AI systems, enhancing their applicability in real-world scenarios and reducing the time and resources required for deployment.

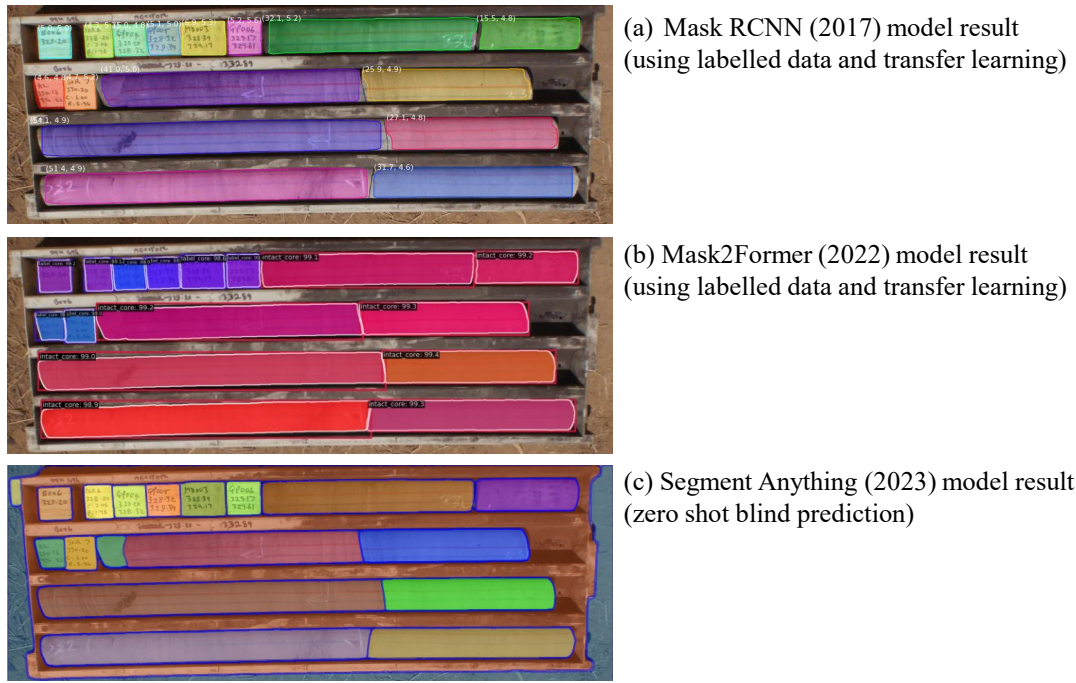

(a) Mask RCNN (2017) model result
(using labelled data and transfer learning)

(b) Mask2Former (2022) model result
(using labelled data and transfer learning)

(c) Segment Anything (2023) model result
(zero shot blind prediction)

**Figure 12.** Comparative analysis of core sample instance segmentation (masks over image) using three AI models: Mask R-CNN, Mask2Former, and Segment Anything (SAM), highlighting their respective accuracies in core sample identification.

### 3.8. Zero-shot learning on digital rock physics

The need for high-accuracy segmentation in reservoir modelling has also been met with the powerful, zero-shot learning SAM model, demonstrating an ability to segment images interactively and automatically without extensive labelled datasets. Traditional semantic segmentation models relied heavily on large-scale annotated datasets, making them time-consuming to prepare, especially for complex CT and SEM rock images. SAM improves the segmentation process by delivering a zero-shot segmentation capability, which is important for digital rock physics with its limited data and intricate image features. Ma et al., (2023) has addressed the zero-shot digital rock image segmentation with a fine-tuned SAM model. Initially, the model encodes images and prompts into high-dimensional vectors using a Vision Transformer and a text encoder, respectively. These vectors are then merged and processed by a mask decoder to produce precise segmentation masks corresponding to the prompts. The versatility of SAM is highlighted through its support for various prompt types—points, boxes, or text—each facilitating different segmentation strategies. Text prompts allow for the segmentation of objects based on abstract concepts, while points and boxes guide the model spatially, enhancing the accuracy of the segmentation process. The SAM model, however, faced limitations when applied to digital rock images due to their complexity and low-contrast features. To address this, the model underwent fine-tuning, resulting in the RockSAM variant, which improved segmentation accuracy for digital rock images without sacrificing its zero-shot learning capabilities (**Figure 13**). This adaptation ensures RockSAM's effectiveness even on resource-constrained devices, presenting a valuable tool for digital rock image analysis.

The RockSAM has shown remarkable efficiency in generating high-quality segmentation masks, overcoming the need for detailed training or complex labelling. Its generalized approach and minimal data requirements stand in contrast to previous models, which, despite their advancements, still needed labelled data and extensive transfer learning. Thus, RockSAM represents a more evolved AI that aligns with the principles of AGI—learning and adapting with minimal human intervention and data. It not only elevates the accuracy of digital rock image analysis but also suggests a future where AGI can seamlessly manage other complex tasks in the oil and gas sector. This trend towards AGI-equipped models suggests an upcoming era of more autonomous, efficient, and versatile AI applications in the industry. The success of RockSAM advances this movement, demonstrating that even with the complex and nuanced task of rock image segmentation, AI can learn to excel with limited datasets. As AI continues to advance, we anticipate models becoming even more adept at handling the challenges and complexities of the natural resource sector.

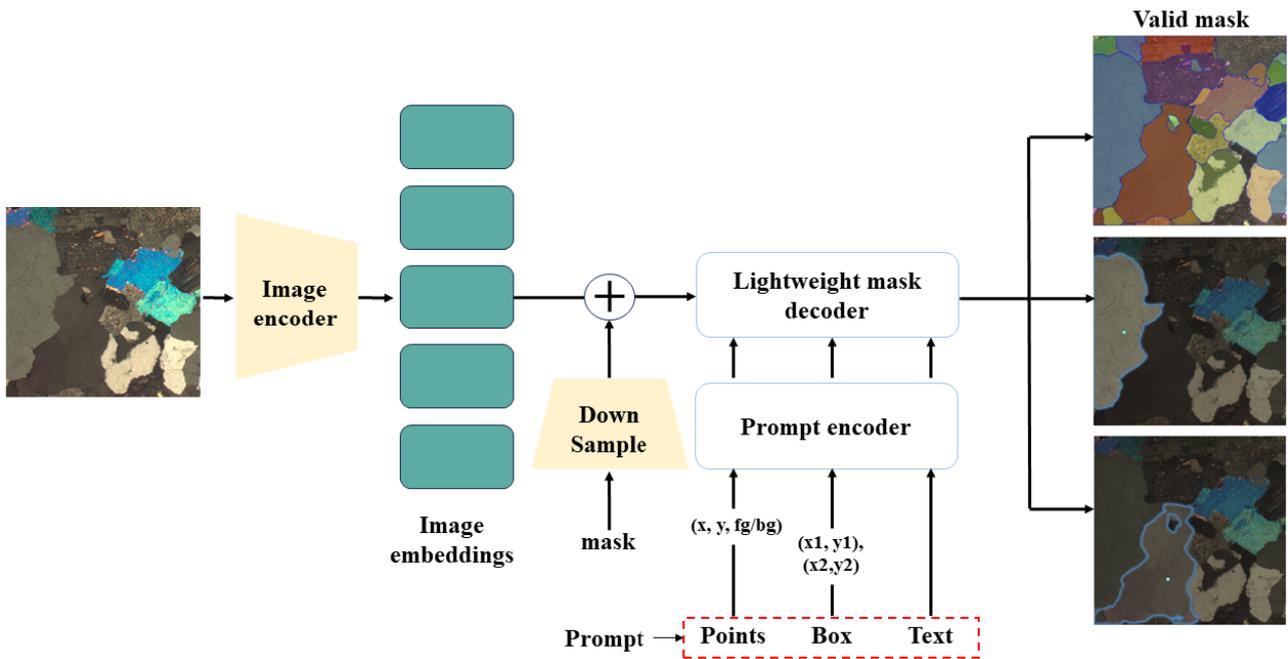

Figure 13. RockSAM, fine-tuned the SAM model for digital rock images, modified the weights in the mask decoder, while maintaining the other components as they are. This process initiates with the transformation of the image into a high-dimensional vector, capturing its intricate details. The prompt (points, a box, or text) is converted into a unique vector and merged into the mask decoder (After (Ma et al., 2023)).

### 3.9. Summary

This section discusses the implementations of AGI within the oil and gas industry, highlighting the transition from conventional operational frameworks to more intelligent, efficient, and sustainable practices. The section reviews key applications of large language models (LLMs), LLM-based agents, and multimodal systems, underscoring the pathways toward realizing an AGI application in the industry. It demonstrates how adopting AGI can bring about comprehensive cognitive functions, enabling systems to perform a wide range of tasks autonomously. This includes optimizing operations and reducing environmental impact.

Specific advancements and initiatives are outlined, including the strategic finetuning of LLMs on domain knowledge, enhancing equipment maintenance through AI, and using multimodal AI to improve real-time production prediction. Each of these points is discussed with examples and case studies that illustrate the potential benefits and challenges associated with integrating AGI technologies into various facets of the oil and gas industry.

### 4. Challenges and future perspectives: shift towards agent-oriented models

Large Language Models (LLMs) and their multimodal form advanced with vision perception capability excel in processing and responding to complex queries, thanks to their sophisticated reasoning and computer vision capabilities. LLMs are adept at breaking down intricate problems into manageable steps, providing comprehensive solutions, actions, and evaluations step-by-step. This capability enables them to address a broad spectrum of challenges effectively.

However, despite these capabilities, LLMs are inherently limited as self-contained systems. They cannot access up-to-date data or draw upon specialized knowledge specific to particular fields, which can result in the production of inaccurate information or "hallucinations." Traditional methods of fine-tuning pre-trained LLMs involve specific adjustments to the neural network weights and datasets, which may restrict their general applicability. Furthermore, LLMs typically face challenges with tasks that require precise numerical computations or staying updated with the latest developments.

To mitigate these limitations, there is a growing recognition of the necessity to integrate external data sources and additional tools. Thus, developing autonomous agents that work in conjunction with LLMs is becoming increasingly crucial (Xi et al., 2023). These agents enhance the functionality of LLMs by enabling real-time data integration and application-specific adaptability (Guo et al., 2024; Wang et al., 2024).

Due to the reasons stated above, since the second half of 2023, there has been a redirection in the strategic focus of leading technology firms and academics, with a move towards the development and implementation of agent-based systems. OpenAI API allows the user to call the core LLM functions to build their own smart agent, reflecting a broader industry trend (GPT-5, 2023). Concurrently, Amazon announced the launch of Amazon Bedrock Agents, augmenting their AI capabilities with advanced agent functionalities designed to handle more sophisticated tasks autonomously (AWS, 2023). Google developed the Vertex AI Agent Builder, which allows developers build and deploy gen AI experiences using natural language or open-source frameworks like LangChain (Google, 2024). These developments are indicative of an industry-wide toward agent-based applications, which was presaged by earlier projects such as Stanford's "AI Town," demonstrating the practical integration of AI agents in simulated environments (Park et al., 2023). Liu et al., (2023b) released AgentBench, an AI agent capability evaluation tool, which provides a standardized framework for assessing the performance and capabilities of AI agents. Ruan et al., (2023) proposes a structured framework for LLM-based AI agents, introduces two types of agents for executing tasks, and evaluates their abilities in Task Planning and Tool Usage (TPTU), aiming to enhance the application of LLMs in complex real-world scenarios and identify areas for further research and development.

All these technological advancements and the shift towards AI agents would hold profound implications for the oil and gas industry. These agents offer the potential to revolutionize industry operations by enabling real-time data integration, automating complex decision-making processes, and potentially overseeing complete operational workflows autonomously. The alignment of these advanced agents with the industry's digital transformation goals suggests a future where AI agents not only augment but centrally manage critical industry operations. As depicted in **Figure 14**, we would introduce the LLM-based agent structure on a geosteering drilling example, where the AI-driven agent assists in the complex decision-making process of automatic geo-steering, a technology used in directional drilling to navigate through the subsurface to the target zone.

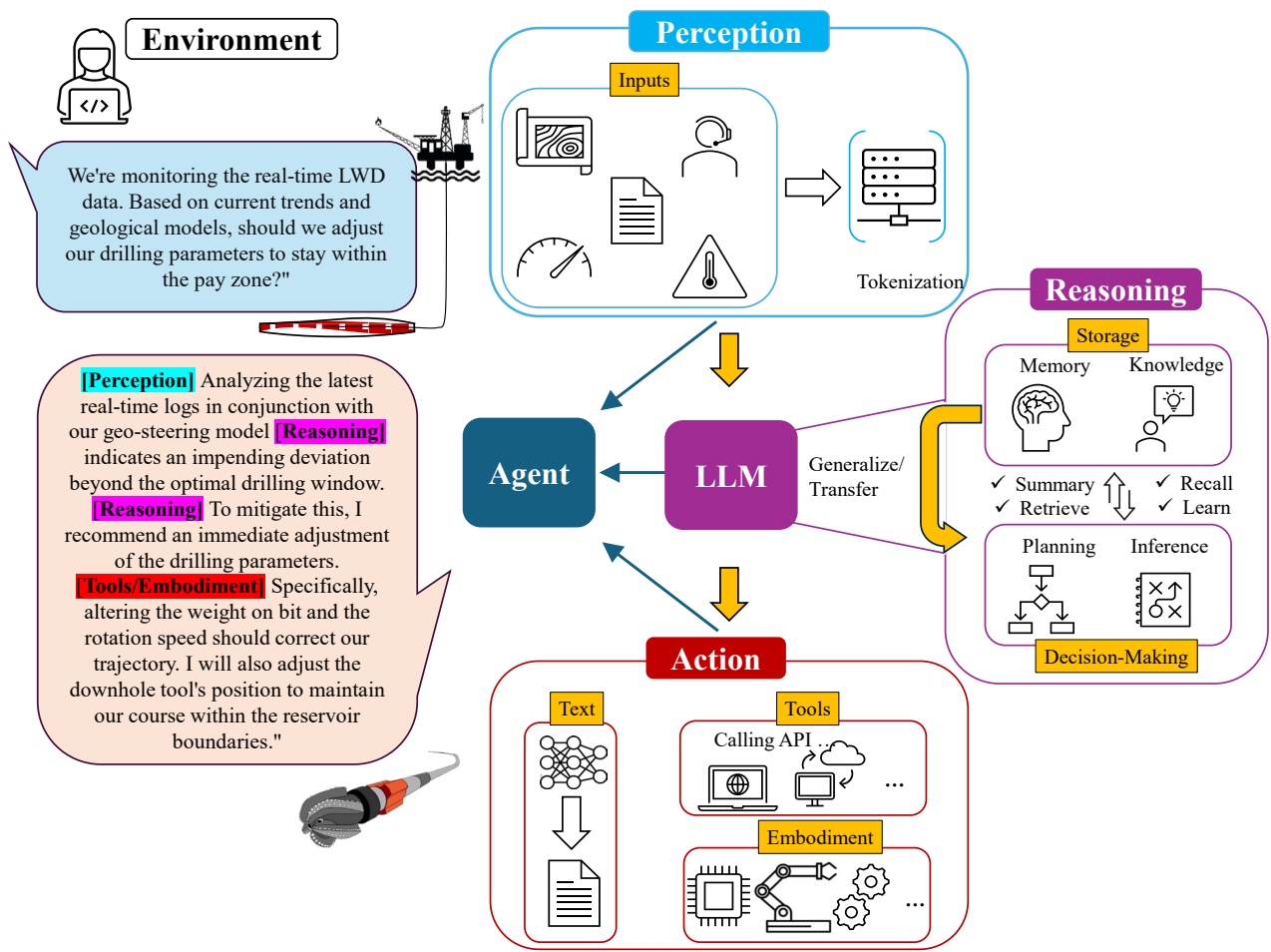

**Figure 14. Enhancing Drilling Precision with LLM-Based Smart Agents in Geo-Steering Operations**

In the context of real-time wellbore positioning, the smart agent's role extends from interpreting live Logging While Drilling (LWD) data to making informed decisions about drill path adjustments. The agent functions as a cognitive intermediary, analysing incoming data streams against geological models to ensure that the drill bit stays within the optimal pay zone. This process is depicted in **Figure 14** through various stages:

(i) **Perception**: The smart agent gathers diverse inputs from environmental sensors, LWD tools, and geological databases, processing this data to maintain an up-to-date perception of the drilling environment.
(ii) **Reasoning:** Leveraging its extensive knowledge base, the agent applies reasoning to understand the implications of the data. For instance, it can predict the trajectory of the drill bit and anticipate the need for adjustments to stay within the reservoir boundaries. The agent also plans actions by considering the current trajectory and the desired outcome. It simulates various scenarios to identify the most effective drilling parameters that align with the geological objectives.

(iii) **Action**: The agent then executes the plan by adjusting the drilling parameters, such as the weight on bit and rotation speed, to correct the drill path, thus embodying the decision-making process in the physical environment of the drilling operation.

The LLM-based smart agent is equipped with advanced cognitive capabilities, simulating a level of understanding and problem-solving that mimics human expertise. The model's ability to interact with text-based prompts enables it to interpret instructions and provide recommendations that are comparable with the expertise of seasoned drillers. The agent's ability to process large volumes of data and make real-time decisions reduces operational downtime and optimizes resource extraction.

Real-world situations frequently require collaborative efforts to achieve optimal task execution. Reflecting the principles of collective human behaviour, researchers have developed a multi-agent framework (Devi et al., 2024; Wu et al., 2023a). This framework is designed to operate collectively, enabling a group of agents to supass the capabilities of individual agents working alone. A multi-agent system (MAS) is a system made up of multiple LLM-based agents that interact with each other in a shared environment (Wu et al., 2023a). These agents are intelligent in the sense that they can make decisions and take actions autonomously, and they can also communicate with each other to achieve a common goal or individual goals that may conflict (Hong et al., 2023; Oroojlooy and Hajinezhad, 2023; Orr and Dutta, 2023). Thus, the MAS are very useful for solving problems that are too complex for a single agent to handle on its own, such as drilling engineering.

The **Figure 15** showcases a network of multiple agents (Agent 1 to Agent 6, each agent has unique expertise besides the general knowledge in the foundation LLM model, such as geology or drilling engineering), highlighting the potential for collaborative decision-making or distributed tasks across various operational sectors. This ensemble illustrates the interconnectedness of the agents, working in unison through complex communication pathways that allow for real-time collaboration and decision-making. The inter-agent dynamics enable the system to adapt dynamically, leveraging the collective intelligence of the group to solve problems more efficiently than any agent could independently.

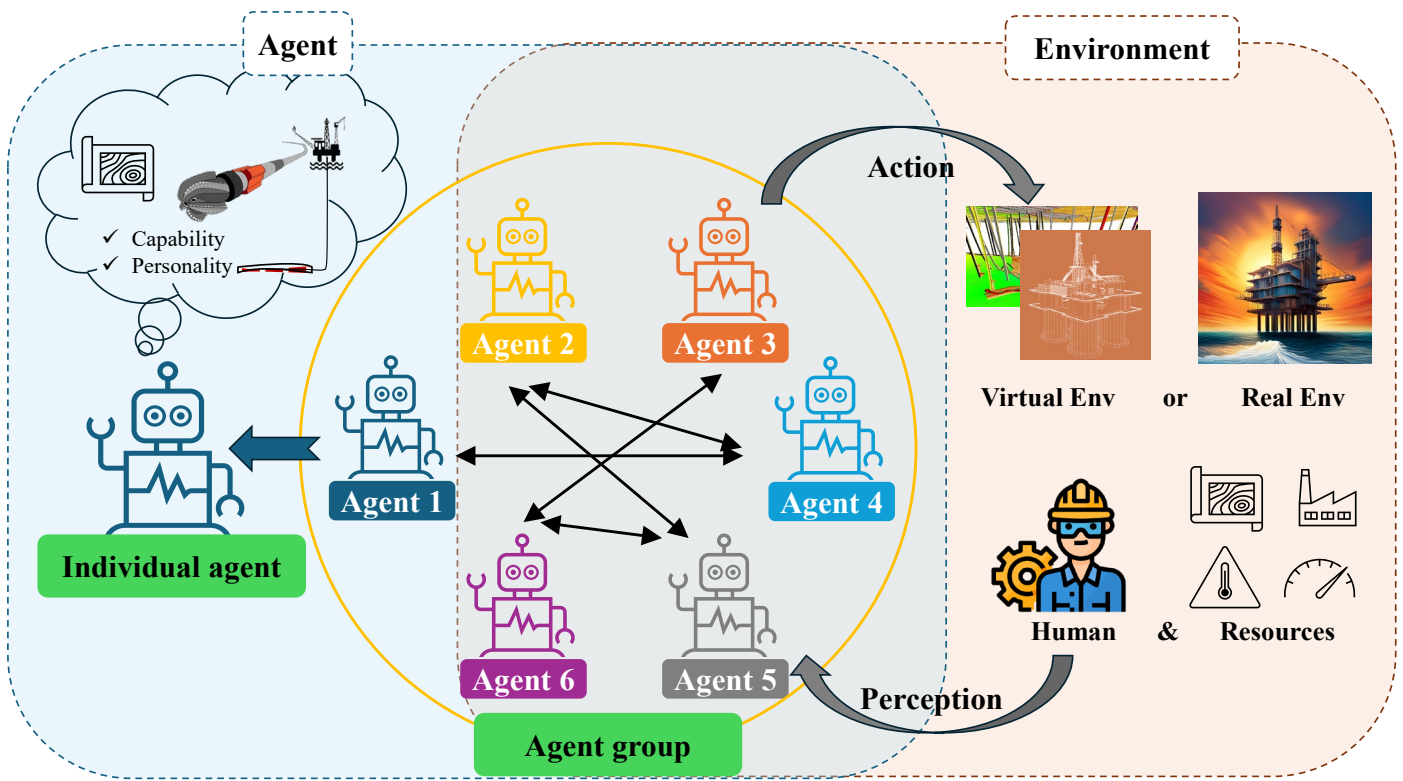

**Figure 15. Multi-Agent Collaboration Framework Enhancing Geo-Steering Drilling Operations (Modified from Xi et al., (2023))**

The agents interact with the "Environment," which is illustrated in two domains: the virtual (e.g., subsurface geological model and well plan) and the real (e.g., hardware, tools on a drilling rig). This dual representation indicates the system's flexibility in operating within simulated models—useful for planning and forecasting—as well as its capacity to act in the physical world, directly influencing drilling operations on an oil rig. Actions taken by the agents are reflected in the environment, influencing factors such as drilling trajectory and equipment adjustments. Simultaneously, the agents receive feedback from the environment ("Reception"), which they use to refine their future actions, thus creating a feedback loop that enhances the operation's precision. The "Human" and "Resources" within the environment suggests the integration of human expertise and available materials and data into the decision-making process. This human intervention ensures that the automated system remains aligned with human oversight and industry standards, optimizing the use of available resources.

## 5. Conclusions

The integration of Artificial General Intelligence (AGI) in the oil and gas industry marks the beginning of a transformative era characterized by the deployment of intelligent systems capable of performing complex tasks with near-human intelligence. This review paper has underscored the strategic importance of AGI's innovative applications and its potential to tackle operational challenges within the upstream sectors of the industry. Through Large Language Models (LLMs) and advanced computer vision systems, AGI has already started to enhance exploration rates, drilling efficiency, safety, and sustainability.

Despite the promising future AGI signals, the industry faces a lot of challenges in adopting these technologies on a large scale. Among these are the needs for skilled professionals for effective implementation, limitations of model training on constrained datasets affecting adaptability, and the essential integration of domain-specific knowledge for maximizing the benefits of AGI technologies. As the technology progress, the move towards agent-oriented models (LLM-based agent) presents an opportunity to redefine the landscape of industry operations. The smart agents, built on the foundations of AGI, are set to play an important role in automating decision-making processes, reducing operational downtimes, and optimizing resource extraction. The successful application of the Segment Anything Model (SAM) in zero-shot learning tasks, such as drill core image analysis and digital rock physics, demonstrates the strides AGI has made in achieving operational efficiency.

The trajectory towards AGI is clear, with a focus on refining the collaborative and autonomous functions of multi-agent systems. These systems promise enhanced precision in complex environments, like those encountered in geosteering drilling operations, where real-time data interpretation and adaptive decision-making are crucial.

This review has articulated the vast potential AGI holds and the pathway toward its realization in the oil and gas industry. As the technology continues to advance, it is expected to bring forth an era of more autonomous, efficient, and versatile AI applications, driving innovation and productivity in the sector.

## 6. Declaration of the use of generative AI

The authors acknowledge the use of ChatGPT, a large language model developed by OpenAI, and Grammarly, a writing and grammar-checking software, in preparing this article. They were used to improve the writing quality of this paper.